**Hierarchically Arranged Helical Fiber Actuators Derived from Commercial Cloth**

By *Jiang Gong,[1] Huijuan Lin,[1] John W. C. Dunlop[2] and Jiayin Yuan*[1]*

[1]Dr. J. Gong, H. Lin, Dr. J. Yuan (Corresponding-Author)
Department of Colloid Chemistry, Max Planck Institute of Colloids and Interfaces, Research Campus Golm, D-14476, Potsdam, Germany (E-mail: jiayin.yuan@mpikg.mpg.de)
[2]Dr. J. W. C. Dunlop
Department of Biomaterials, Max Planck Institute of Colloids and Interfaces, Research Campus Golm, D-14476, Potsdam, Germany

The development of advanced materials carrying implemented stimuli-responsiveness and reversible actuation is actively pursued in the fields ranging from medicine and biology to materials science, chemistry, physics and engineering.[1-4] Their functions to alter volume and/or shape can be triggered by environmental variations,[5-8] such as water (or humidity),[9,10] electricity,[11-13] and light.[14] Their diverse applications include artificial muscles,[5,15] sensors,[16] electric generators,[9,17] switches,[18] robotics,[19-21] "smart" materials,[10,13,22,23] *etc*. Significant progress has been achieved over the past decade in preparing artificial actuators from carbon nanotubes (CNTs),[24-26] graphene,[27] and more.[18,28,29] Currently, it remains a prosperous field full of challenges, in which modern design concepts are eagerly desired to meet scientific curiosity and practical usage.

Numerous responsive materials can be found in nature that inspire materials scientists to exploit their underlying mechanisms to create innovative biomimetic materials and actuating devices.[30-33] Fascinating examples include the opening of pine cones,[34] arching upward of spruce branches,[35] unfolding of seed capsules,[36] and twisting of seed pods.[37] Two fundamental actuation forms—bending and twisting—are recognized.[35,37,38] Bending is achievable by a bilayer configuration in which opposing tissues possess different cellulose fibril orientations in the cell walls, while in isolated cells with helically arranged fibrils, the contraction of matrix leads to a twisting motion.[39,40] Learning from nature, environment-responsive devices that perform similarly or superiorly to their nature counterparts were fabricated.[41-43] Most of these efforts however suffer from drawbacks such as complicated synthesis, small size-scale deformation, low productivity, and/or poor mechanical properties.

Because of low-cost, rich abundance, lightweight and high mechanical flexibility, cloth made from cotton has followed human history for thousands of years and nowadays remains indispensable in our modern life. Inspired by helical microfibrils in plant cells to tailor mechanical deformation, for the first time we here report a unique example of novel hygroscopic cloth actuator that offers diverse motion forms, multiple functions, and a scale-up of production. The actuator is structurally composed of a scaffold made up of a commercial cloth that contains perpendicularly arranged helical cotton microfiber arrays and polyamide microfiber bundles (termed "native cloth"), and an *in situ* generated nanoporous polymer hybrid inside the scaffold. The former provides multiple forms of actuations including folding, twisting and rolling, while the latter serves as a function amplifier to boost its actuation performance, *e.g.*, in an electric generator, in which the yielded voltage is two orders of magnitude of the state-of-the-art actuators of similar types.

**Figure 1** schematically illustrates the process to fabricate the targeted cloth actuator. Initially, a mixture solution of a cationic poly(ionic liquid) (structure and characterization shown in Figure S1−S3) and poly(acrylic acid) at a 1:1 molar ratio of the monomer unit was prepared in *N,N*-dimethylformamide (DMF). Subsequently, a defined amount of CNTs (Figure S4) was added under sonication to reach a homogeneous polymer/CNT dispersion. The dispersion was cast onto a flat native cloth of a defined size on a clean glass plate. After drying, the formed polymer/CNT/cloth hybrid (termed "actuator intermediate"), which adhered firmly to the glass plate, was soaked in a 0.5 wt % of aqueous NH$_3$ solution. The ammonia treatment here is a necessary step to convert the freshly introduced polymer/CNT blend in the cloth into a 3D interconnected nanoporous network.[8,44] Finally, a mechanically flexible, free-standing, nanoporous cloth actuator was easily peeled off from the glass substrate.

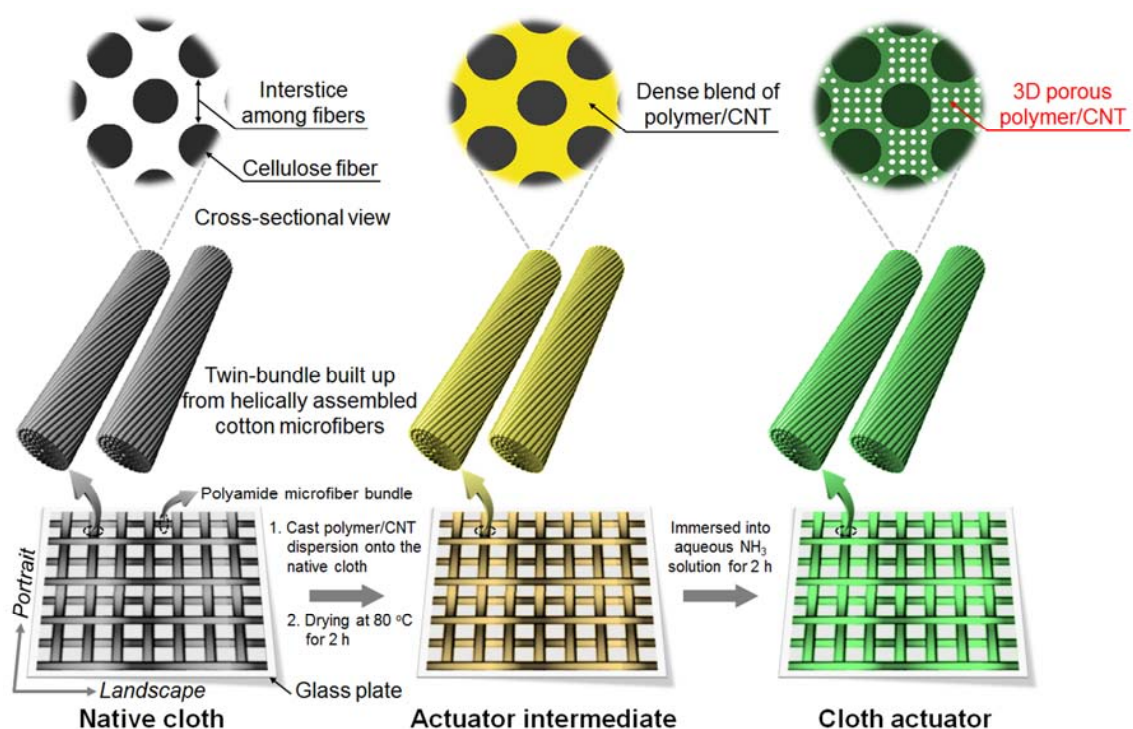

**Figure 1**. Fabrication scheme of the cloth actuator from a native cloth (containing helical cotton microfiber arrays in the portrait orientation and polyamide microfiber bundles in the landscape orientation) and an *in situ* generated 3D porous polymer/CNT hybrid network.

Scanning electron microscopy (SEM) details the structural evolution from the native cloth to cloth actuator. The native cloth displays a hierarchical, rectangular network with a grid size of ca. 1.3 mm × 1.6 mm containing bundles of cellulose fibers in the portrait orientation and bundles of polyamide fibers in the landscape orientation (**Figure 2**a,S5,S6). Every two bundles of cellulose fibers are twinned to construct a single grid line (Figure 2b). Each bundle comprises tens of individual cellulose fibers of 4–10 μm in diameter. Uniquely, these fibers are oblique to the longitudinal axis of bundles at an angle of 10–15º. This helical ordering of fibers in each bundle is neither a nature product nor home-made but occurs in the factory as a necessary processing step to minimize the sliding in fibers. The front and cross-sectional views of fibers in Figure 2c and S7, respectively, reveal the relatively smooth surface and interstitial space among them, which can accommodate the to-be-introduced porous polymer/CNT hybrid.

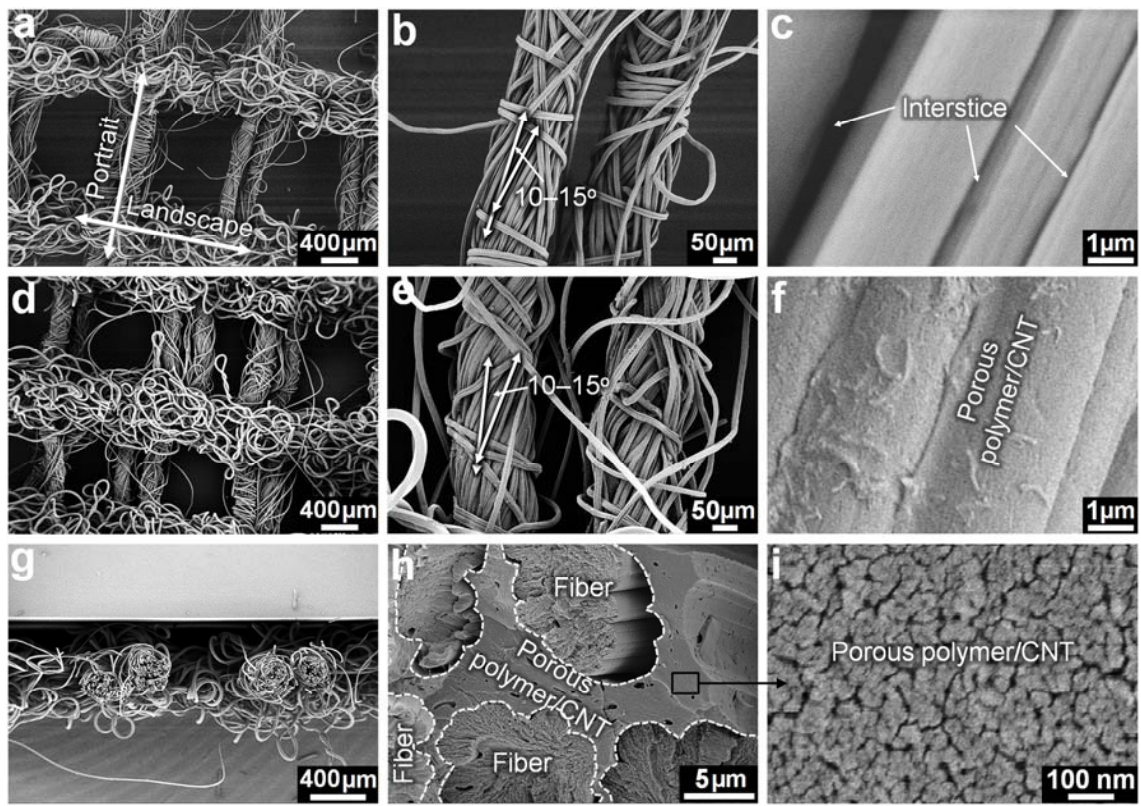

**Figure 2.** SEM images of the native cloth received from the factory (a–c) and cloth actuator (d–f for front view and g–i for cross-sectional view).

The interstice among fibers within individual bundles in the actuator intermediate is no longer visible by SEM observation (Figure S8), which results from the infiltration of the polymer/CNT dispersion into the microvoids. Meanwhile the milimeter sized grid windows remain open. No nanopores are found at this stage (Figure S8). After ammonia treatment, the fabricated cloth actuator preserves the general micron-scale structure of the native cloth (Figure 2d–2i). By contrast, on a nanoscale pores of 20–40 nm in size are observed in the cross-sectional view of individual bundles (Figure 2i). The pore size is slightly smaller than that of cloth actuator prepared in the absence of CNTs (Figure S9–S11). It should be mentioned that a particular feature of our actuator design is the employment of commercial cloth, in which the helical configuration of cellulose microfibers, an important yet time-demanding synthetic step, was conducted industrially thus saving tremendous labour in laboratory. They are above all cheap (~1 Euro kg$^{-1}$), and readily scalable in production. As a proof of concept, a flexible cloth actuator of 6400 cm$^2$ in size is prepared (Figure S12).

A striking feature of our cloth actuator is the tunable actuation merely *via* controlling the aspect ratio without altering other properties. As shown in **Figure 3**a–c, upon drying in an atmospheric environment (20 $^o$C, 70% RH), a single twin-bundle actuator (1.3 mm × 80 mm) folds into a distorted thread (Figure S13, Movie S1), while a double twin-bundle actuator (2.5 mm × 80 mm) twists into a DNA-like helix, similar to a triple twin-bundle actuator (3.8 mm × 80 mm) that forms a triple-helix line (Movie S2). The helix is also observed in actuators carrying up to 8 twin-bundles, after which the cloth actuator starts a helix-to-tube transition. A flat cloth actuator containing 16 twin-bundles (20 mm × 80 mm) rolls into a cylindrical tube with an inner diameter of 4 mm (Figure 3d,S14, Movie S3). Upon contact with water, all of these different shapes rapidly recover their original forms (Movie S4–S7). Thus, unlike the previous report,[45] tunable actuations of folding, twisting and rolling are readily achieved in our actuator design from the same material without complex synthetic procedure and/or sophisticated equipment. Other shapes of the cloth actuator are also created to demonstrate the complex actuation forms (Figure S15). Besides, to quantify the actuation behavior, revolution number is used here. As depicted in Figure 3e, with increasing twin-bundle number in the actuator, the revolution number decreases, *e.g.*, from 11 for the double twin-bundle actuator to 3 for the flat actuator bearing 16 twin-bundles.

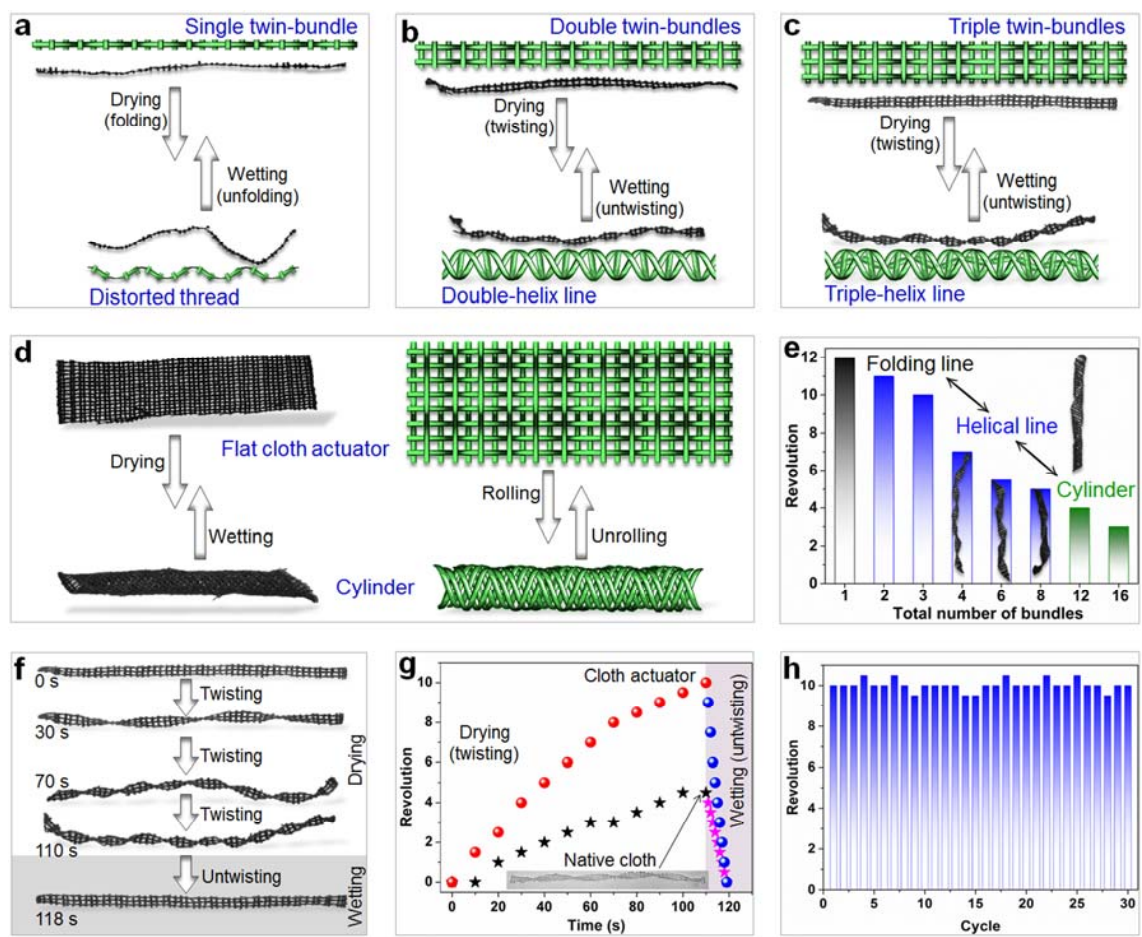

**Figure 3.** Schematic illustration and adaptive movement of (a) single twin-bundle actuator, (b) double twin-bundle actuator, (c) triple twin-bundle actuator, and (d) flat cloth actuator bearing 16 twin-bundles. (e) Plot of the revolution number *vs.* number of bundles; the insets show photographs of cloth actuators after drying. (f) Adaptive movement upon time for the triple twin-bundle actuator. (g) Plot of revolution number *vs.* time for the triple twin-bundle actuator and the native cloth. (h) Cycling test of the triple twin-bundle actuator during periodic drying-wetting cycles.

The actuation of helically arranged microfibers/nanotubes in bundles has been reported previously to be induced by collective expansion/contraction of individual microfibers/nanotubes.[9,11,46] The adaptive movement upon time in a triple twin-bundle actuator was observed as an example to clarify the actuation behavior of cloth actuator during drying-wetting (Figure 3f). Upon drying, water molecules in the cloth actuator are continuously lost, resulting in the contraction of the helically arranged microfibers. An obvious twisting is observed in the initial 30 s, forming a triple-helix line with a revolution number of 4. The twisting movement continues as time elapses. It takes ca. 80 s for the actuator to proceed from an obvious twisting movement to a dense triple-helix with a revolution number of 10. This value is twice that of native cloth (Figure 3g). Apparently, the incorporation of the nanoporous polymer/CNT hybrid into the cloth scaffold enhances the actuation performance. In control experiments, the nonporous actuator intermediate and CNT-free cloth actuator are observed to carry a revolution number of only 5 and 7, respectively, which are higher than native cloth but obviously lower than the carefully fabricated cloth actuator. These results prove that the presence of 3D nanoporous network and CNTs in the bundles indeed promotes the actuation function. Besides, the actuation is reproducible (Figure S16–S18), reversible and repeatable, as exemplified here with 30 cycles (Figure 3h,S19). It also shows high stability upon storage in water for at least 3 months (Figure S19).

Our fiber actuator design is assumed to be closely correlated to two structure features, that is, the unique spatial packing order of microfibers and the engineered porous matrix (Figure S20). It has been demonstrated previously that the nano-/micropores of hierarchically helical CNT fibers are beneficial to the diffusion of solvent among fibers.[46] We find that the pore size of the porous polymer hybrid, when below 30 nm, affects the actuation performance (Figure S21). Equally important, the 3D nanoporous polymer/CNT hybrid connects and holds closely the packed microfibers together as a single entity to cancel relative sliding among microfibers to avoid energy dissipation due to inter-fiber friction. The introduction of porous networks thus seems an effective way to fabricate high-performance actuators.

The mechanical property of actuators is of paramount importance for practical applications. The tensile strength and elongation at break of cloth actuator are up to 70.4 MPa and 55.3%, respectively, which are favorable properties stemming from the native cloth substrate (Figure S22,S23). The tensile strength of cloth actuator is close to that of graphene oxide fiber actuator[9] and robust polypyrrole composite actuator,[41] but its elongation at break is 4 and 2 times greater, respectively. Furthermore, the cloth actuator still works well even after the application of moderate tensile strain (below 30%, Figure S24) or harsh washing treatments such as vigorous agitation and sonication in solution (Figure S25).

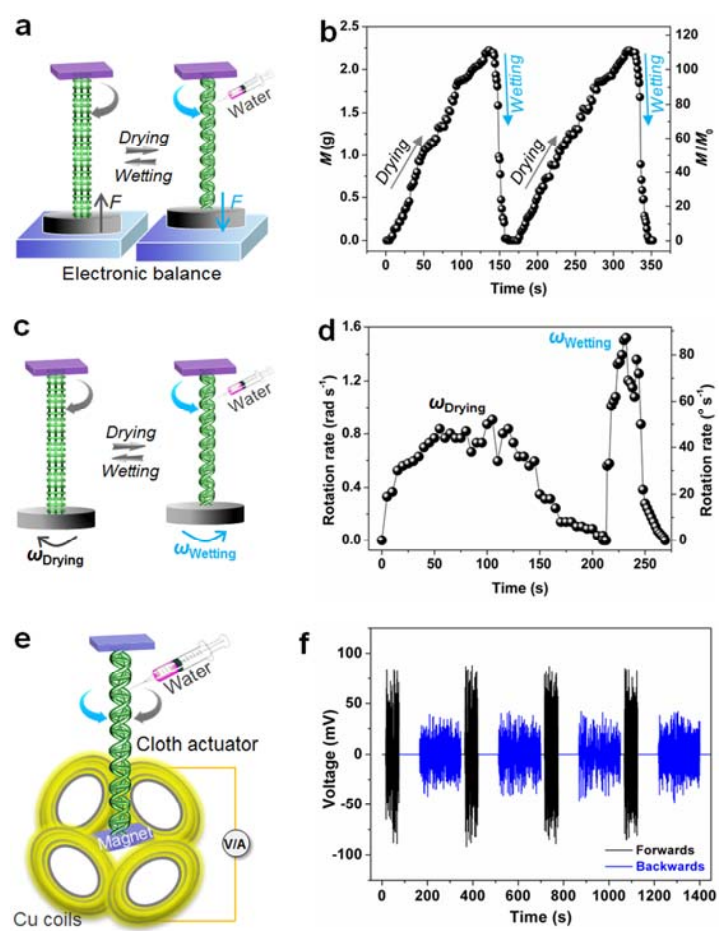

**Figure 4.** (a) Schematic illustration of the experimental setup to measure force generated by cloth actuator in the portrait orientation. (b) Its force *vs.* time plot; *M* is the force exerted on the balance by cloth actuator; $M_0$ is the weight of cloth actuator. (c) Schematic illustration of the experimental setup to quantitatively measure the rotation. (d) Its rotation rate *vs.* time plot. (e) Schematic illustration of an electric generator. (f) Plot of the open-circuit voltage produced by actuator rotation (forwards by wetting and backwards by drying) *vs.* time.

Next, we quantified the mechanical force generated by the twisting of a triple twin-bundle actuator (3.8 mm × 80 mm, ~20 mg). One of its ends in the portrait orientation was tethered to a reference sample on an electronic balance (**Figure 4**a). During drying it pulls up the reference and simultaneously decreases the pressing force that the reference exerts on the balance, and *vice versa* during wetting. The contraction and stretching forces generated by the actuator upon alternation of twisting and untwisting, which are read out from the balance in the form of weight, are recorded (Figure 4b). The maximum force detected is 2.2 g, which is 110 times of its own weight and 4 times that of native cloth (Figure S26). The forces generated by the single twin-bundle actuator and flat cloth actuator are also measured (Figure S27,S28). To quantify the rotation actuation, a disc-shaped reference was tethered to one end of the cloth actuator (3.8 mm × 160 mm) (Figure 4c). A rotation of 34 revolutions (*i.e.*, 212 revolutions m$^{-1}$) is observed upon a dry-wetting cycle even though the reference mass is 25 times that of the actuator. The initial reference acceleration ($\alpha$) upon wetting in Figure 4d is calculated to be 0.28 rad s$^{-2}$. Since the moment of inertia of the reference ($J$) is 4.5 × 10$^{-7}$ kg m$^{-2}$, the maximum start-up torque ($\tau$) is $\tau = J\alpha = 1.26 \times 10^{-7}$ N m$^{-1}$, which is 7 times that of native cloth (Figure S29) and comparable to carbon-based actuators.[9,46]

The reversible, tunable actuations of the cloth actuator endow it with a variety of fascinating applications. Recently, water-powered electric generator has received extensive attention because of its environmentally friendly nature.[17,47] By virtue of the rotary actuation in response to water, a cloth actuator-based electric generator is fabricated (Figure 4e,S30).

Wetted by water periodically, the cloth actuator drives reversible rotation of a magnet surrounded by copper coils to generate electric current. An open-circuit voltage of 75 mV is produced (Figure 4f), which largely exceeds a similar but carbon-based design model reported by others (1 mV),[9] and is 3 times that of the native cloth-based generator (Figure S31). Furthermore, the rolling function of cloth actuator, if necessary, can be applied as, for instance, a mechanical gripper and a "smart" window (Figure S32,S33). Additionally, the recent progress in complex 3D deformations[48,49] motivates us to expand the designs of cloth actuators for more sophisticated applications in the next stage of our work.

In summary, inspired by the helical microfibrils in plant cells, we have developed an unusual hygroscopic cloth actuator *via* modification of commercial cloth that bears already helically arranged microfibers. The remarkably improved actuation performance plus large-scale accessibility and favorable mechanical flexibility of the cloth templates indicates the success of a new general actuator design principle discussed here for multifunctional tasks. It is anticipated that without complicated, time-consuming synthetic efforts, a variety of advanced actuators can be easily accessed *via* adopting available and appropriate templates, which exists in a large amount in either nature or our daily life.

*Acknowledgements*
This work was financially supported by the Max Planck Society and the European Research Council (ERC) Starting Grant with Project No. 639720–NAPOLI.

Supporting Information

**Hierarchically Arranged Helical Fiber Actuators Derived from Commercial Cloth**

*Jiang Gong, Huijuan Lin, John W. C. Dunlop, and Jiayin Yuan\**

*Experimental*

*Materials*: 1-Vinylimidazole (Alfa Aesar, purity ≥ 99%), 2,2'-azobis(2-methylpropionitrile) (AIBN, Sigma-Aldrich, purity ≥ 98%), bis(trifluoromethane sulfonyl)imide lithium salt (LiTf$_2$N, Alfa Aesar, purity ≥ 99%), bromoacetonitrile (Alfa Aesar, purity ≥ 97%), and isophthalic acid (IPA, 166 g mol$^{-1}$, Sigma-Aldrich, purity ≥ 99%) were used as received without further purifications. Dimethyl sulfoxide (DMSO), *N,N*-dimethylformamide (DMF), methanol, and acetate were of analytic grade. Poly(acrylic acid) (PAA, weight-average molecular weight = 1800 and 450,000 g mol$^{-1}$) was purchased from Sigma-Aldrich. The native cloth bearing both cotton and polyamide microfibers was kindly provided by Holthaus Medical GmbH & Co KG (Germany), which shows several advantages over the previously reported actuator materials (*e.g.*, graphene and carbon nanotube (CNTs)). Firstly, it is cheap and displays a high water adsorption capacity (ca. 44 wt %) due to its abundant hydroxyl groups; secondly, when produced, it contains already helically arranged cellulose microfibers in the portrait orientation to improve the frictional force between fibers; thirdly, it shows high tensile strength and good mechanical flexibility. Multi-walled CNTs with a diameter of 35–60 nm and length of ca. 30 μm were purchased from Chengdu Institute of Organic Chemistry (China). CNTs were ball-milled at 400 rpm for 6 h before being used, and the length was reduced to 0.3–1 μm.

*Synthesis of Poly(Ionic Liquid) (PIL)*: As displayed in Figure S1, we firstly synthesized poly(3-cyanomethyl-1-vinylimidazolium bromide) (PCMVImBr) according to our previous method [S1]. Briefly, 10.0 g of the monomer (3-cyanomethyl-1-vinylimidazolium bromide (CMVImBr), prepared from 1-vinylimidazole and bromoacetonitrile), 30 mg of AIBN, and 100 mL of DMSO were loaded into a 200 mL of reactor. The mixture was deoxygenated several times by a freeze-pump-thaw procedure. The reactor was refilled with $N_2$ and placed in an oil bath at 70 $^o$C for 24 h. The mixture was then exhaustively dialyzed against water for 1 week and freeze-dried from water. The gel permeation chromatography (GPC) trace of PCMVImBr is depicted in Figure S2. Poly[3-cyanomethyl-1-vinylimidazolium bis(trifluoromethane sulfonyl)imide] (PCMVImTf$_2$N, simplified as "PIL") was synthesized *via* anion exchange with PCMVImBr using LiTf$_2$N in aqueous solution. The $^1$H nuclear magnetic resonance ($^1$H NMR) spectrum of PIL is displayed in Figure S3.

*Preparation of Cloth Actuator*: 1.00 g of PIL and 0.18 g of PAA (1800 g mol$^{-1}$, the molar ratio of two monomer units is 1:1) were firstly dissolved in 20.0 mL of DMF to form a homogeneous solution. A defined amount of CNTs (morphology characterization shown in Figure S4) was dispersed in the PIL-PAA solution assisted by vigorous agitation for 1 h and sonication for another 1 h to prepare a homogeneous PIL-PAA/CNT dispersion, which was then uniformly cast onto a native cloth of a defined size supported on a clean glass plate. Typically, 0.4 mL of the dispersion containing 20.0 mg of PIL, 3.6 mg of PAA, and 0.7 mg of CNT (the content of CNT is ca. 3 wt % with regard to the total polymer weight) was deposited on a native cloth of 20 mm × 80 mm in size (ca. 83 mg, 20 mm in the landscape orientation and 80 mm in the portrait orientation, as displayed in Figure S5). The obtained composite cloth, termed actuator intermediate, was dried at 80 $^o$C for 2 h and then soaked in a 0.5 wt % of aqueous ammonia solution for 2 h. Afterwards, a mechanically flexible, free-standing hierarchical cloth actuator was easily peeled off from the glass substrate. The weight percentage of the porous polymer/CNT hybrid in the cloth actuator is estimated to be 23 wt %. For comparison, a reference cloth actuator without CNTs was similarly fabricated in the absence of CNTs *via* ammonia treatment. In addition, cotton-free PIL-PAA and PIL-PAA/CNT porous membranes were prepared in the similar way in the absence of the native cloth. Besides, IPA (166 g mol$^{-1}$) and PAA (450,000 g mol$^{-1}$) were used in place of PAA (1800 g mol$^{-1}$) to prepare the related cloth actuators by using the similar approach.

*Characterization*: GPC measurement was conducted at 25 $^{o}$C on a NOVEMA-column with a mixture of 80% acetate buffer and 20% methanol as eluent (flow rate = 1.0 mL min$^{-1}$, PEO standards using RI detector-Optilab-DSP-Interferometric Refractometer). $^{1}$H NMR measurement using DMSO-$d_6$ as solvent was carried out at 25 $^{o}$C using a Bruker DPX-400 spectrometer operating at 400 MHz. Scanning electron microscopy (SEM) measurements were carried out in a LEO 1550-Gemini electron microscope (acceleration voltage = 3 kV). The pore sizes in the porous polymer hybrids were measured using software ImageJ 1.47. Transmission electron microscopy (TEM) measurement was performed using a Zeiss EM 912 (acceleration voltage = 120 kV). Mechanical properties were measured on an Instron 1121 at an extension speed of 10 mm min$^{-1}$. All data were the average of five independent measurements along with the standard error less than 10%. The electro character of electric generator was measured using Digital-Multimeter Benning MM 7-1. The absorbance of CNT aqueous dispersion was measured by UV/Vis/NIR spectrophotometer (Lambda 900).

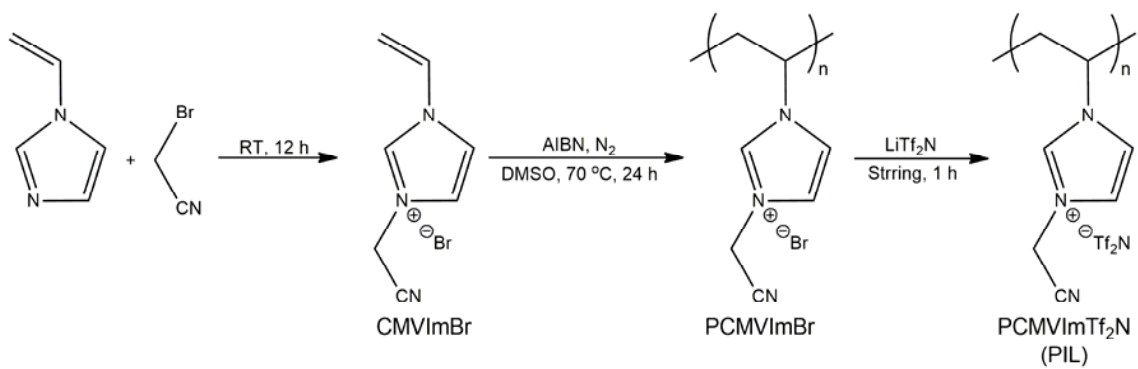

**Figure S1.** Synthesis of a cationic poly(ionic liquid), PCMVImTf$_2$N (abbreviated as "PIL").

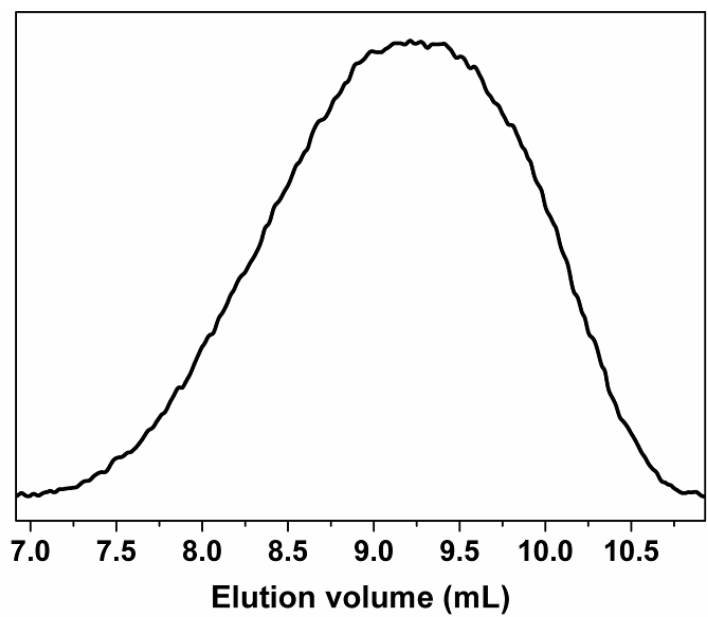

**Figure S2.** GPC trace of PCMVImBr.

The apparent number-average molecular weight and polydispersity index of PCMVImBr are measured to be $1.97 \times 10^5$ g mol$^{-1}$ and 2.67, respectively. PIL used in this study is synthesized by anion exchange of PCMVImBr with LiTf$_2$N in aqueous solution. Therefore, the apparent number-average molecular weight of PIL is calculated to be $3.80 \times 10^5$ g mol$^{-1}$.

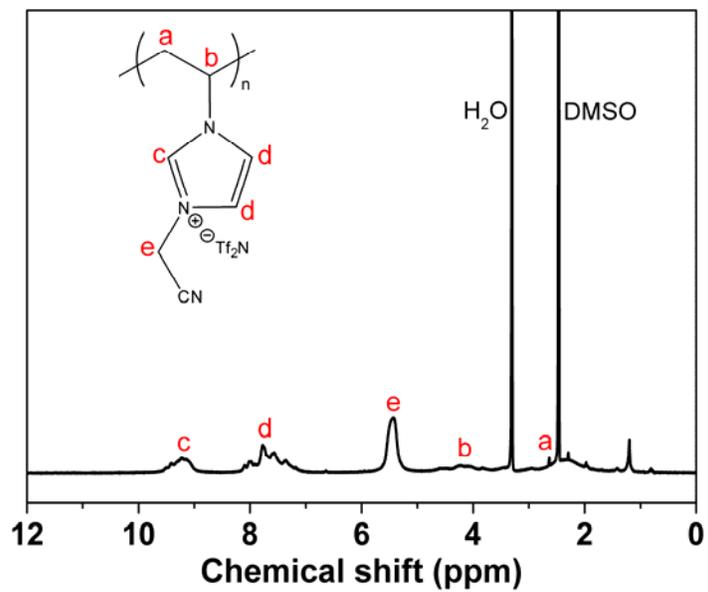

**Figure S3.** ¹H NMR spectrum of PIL using DMSO-$d_6$ as the solvent.

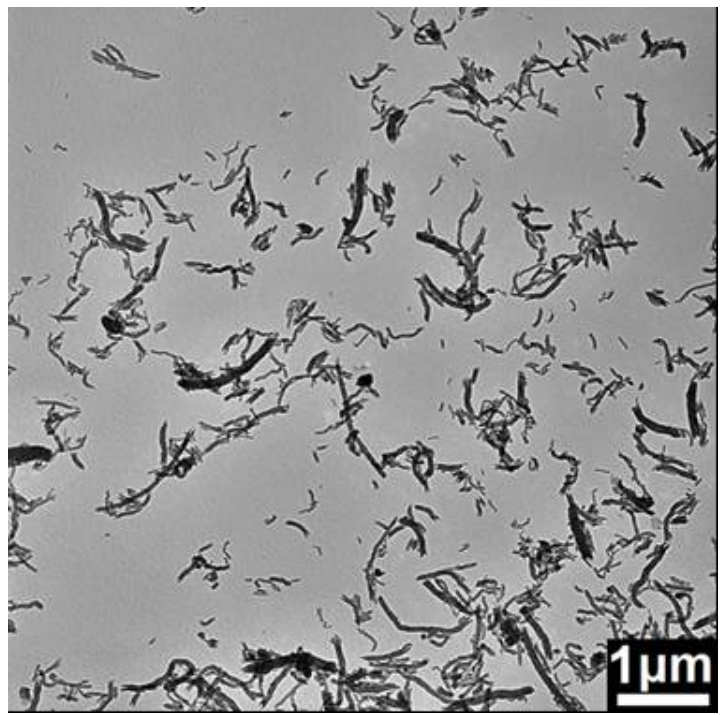

**Figure S4.** TEM image of CNTs with the diameter of 35–60 nm and the length of 0.3–1 μm.

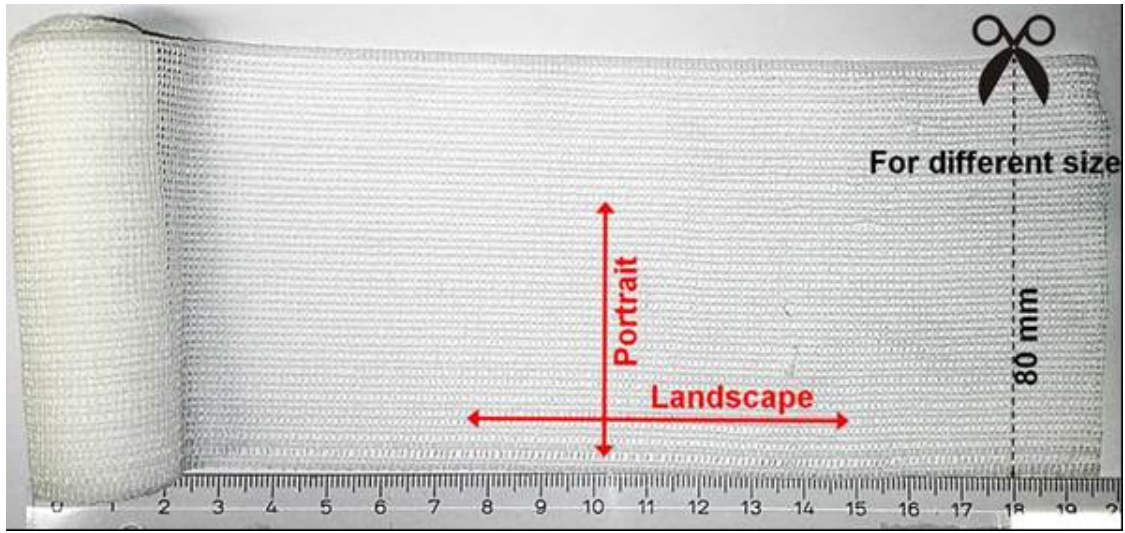

**Figure S5.** Photograph of the native cloth made of the cellulose fiber twin-bundles in the portrait orientation and the polyamide fiber bundles in the landscape orientation according to the information from the supplier.

The native cloth is generally cut in the portrait orientation for preparing the cloth actuator with a fixed length of 80 mm and different width (from 1.3 mm to tens of millimeters). Typically, 16 twin-bundles are equal to 20 mm. Besides, several cloth actuators can be connected in the portrait orientation to prepare a longer cloth actuator, *e.g.*, 160 mm.

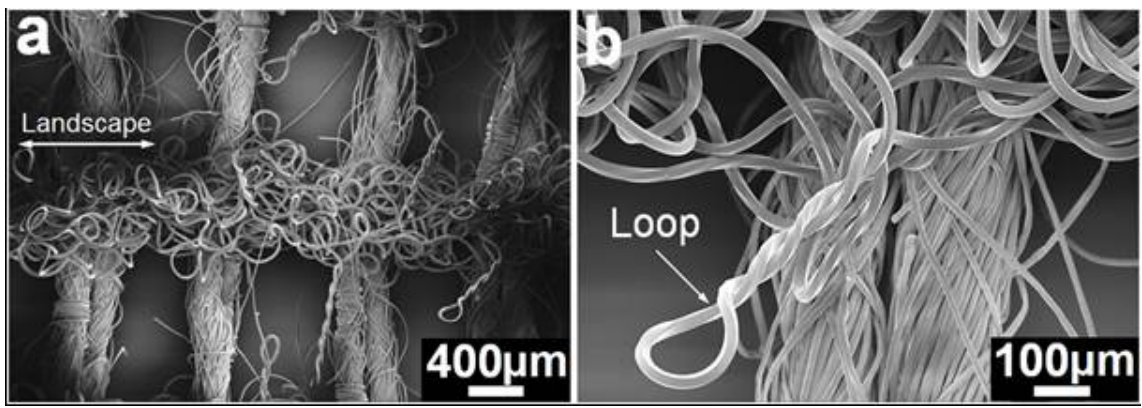

**Figure S6.** SEM images in (a) low and (b) high magnifications of native cloth in landscape orientation, *i.e.*, polyamide fibers (30 wt %, according to the information from supplier).

These polyamide fibers in the landscape orientation can hold cellulose fibers in the portrait orientation due to the formed grid network. Meanwhile, the loose assembly state of polyamide fibers enables the stretchable feature of native cloth fairly well in the portrait orientation. It should be specifically noted that these polyamide fibers are individually twisted in their longitudinal axis to form random loops, as shown in an enlarged view (b). These loops are responsible for the stretching properties for the polyamide fibers in the landscape orientation, that is, when stretched, these polyamide fibers elongate themselves by pulling the loops into fully extended lines. Once the stretching force vanishes, polyamide fibers contract and the loops are re-formed. It was found that when native cloth was coated with the porous polymer/CNT composite, only the bundles of cellulose fibers in the portrait orientation were infiltrated and their interstice was fully filled with the polymer/CNT composite, while the bundles of polyamide fibers in the landscape orientation were hard to be coated by polymer/CNT composite and thus did not support the composite. The influence of polyamide bundles in the landscape orientation on the actuation behavior of the cotton bundles in the portrait orientation is minor and thereby ignored during the discussion, because the actuation performance is dominated by the amplifying effect of the accommodated porous polymer/CNT composite. Additionally, the stretching/contraction of polyamide fibers in response to wetting/dewetting is in the landscape direction, which is perpendicular to cellulose fiber bundles in the portrait orientation and thus does not contribute to the stretching/contraction of cellulose fiber bundles and their twisting.

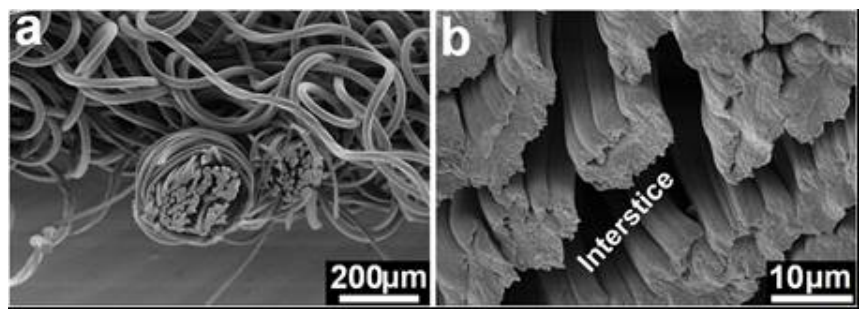

**Figure S7.** SEM images of the cross-sectional structure of the native cloth in the portrait orientation (*i.e.*, cellulose fiber bundles) in (a) low and (b) high magnifications. The interstice can be easily observed among fibers.

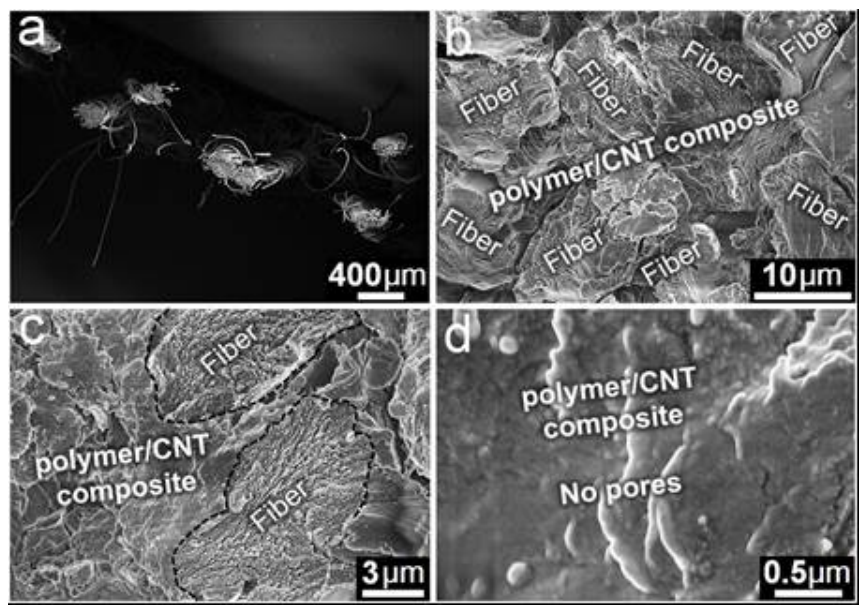

**Figure S8.** SEM images in different magnifications of the cross-sectional structure in the portrait orientation of the nonporous actuator intermediate before ammonium treatment. Experimental condition: the size of native cloth = 20 mm × 80 mm, and the PIL-PAA/CNT dispersion = 0.4 mL containing 20.0 mg of PIL, 3.6 mg of PAA, and 0.7 mg of CNT.

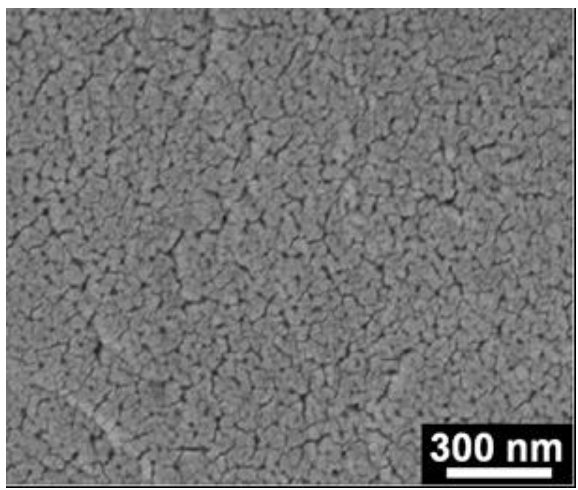

**Figure S9.** SEM image of the nanoporous structure with a size range between 30 and 50 nm of the porous PIL-PAA in the cloth actuator prepared without CNTs.

In comparison to the cloth actuator without CNTs, the slightly decreasing pore size of the porous polymer hybrid in the cloth actuator bearing CNTs is believed to be caused by the presence of CNTs, which to some extent retards the diffusion of $NH_3$ and restricts the phase separation [S2,S3], resulting in a relatively smaller pore size. This speculation is confirmed by the result of the following model experiment as shown in Figure S10,S11.

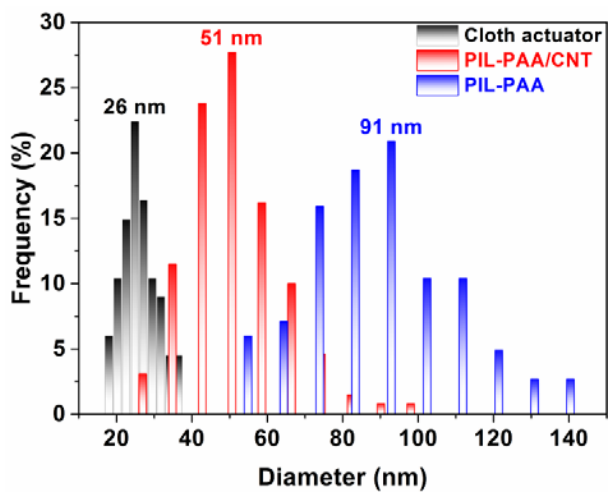

**Figure S10.** Pore size distribution plots of the nanoporous PIL-PAA/CNT hybrid in the cloth actuator and the cotton-free PIL-PAA/CNT porous membrane with 3 wt % CNT addition, and the nanoporous PIL-PAA hybrid in the CNT-free, cotton-free PIL-PAA porous membrane.

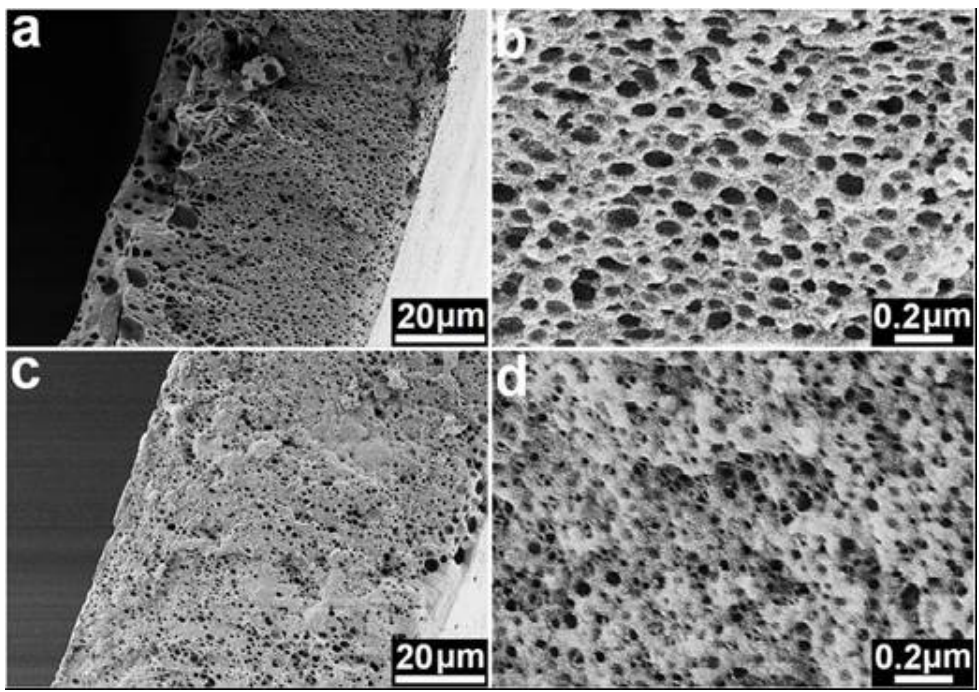

**Figure S11.** SEM images of (a and b) the cotton-free, CNT-free PIL-PAA porous membrane, and (c and d) the cotton-free PIL-PAA/CNT porous membrane with 3 wt % CNT addition.

The cotton-free PIL-PAA/CNT porous membrane shows a smaller pore size than the cotton-free, CNT-free PIL-PAA porous membrane, indicating that the presence of CNTs indeed decreases the pore size of the porous polymer.

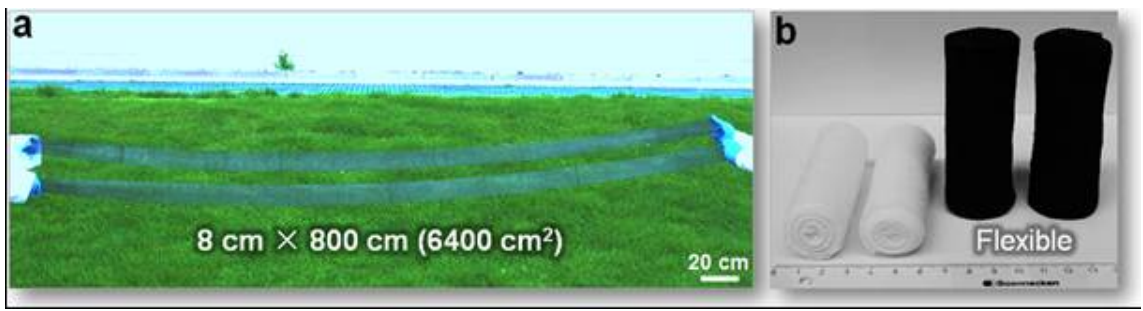

**Figure S12.** Photographs of (a) the cloth actuator with a size of 8 cm × 800 cm (6400 cm$^2$), (b, left) the native cloth (white), and (b, right) the flexible cloth actuator (black).

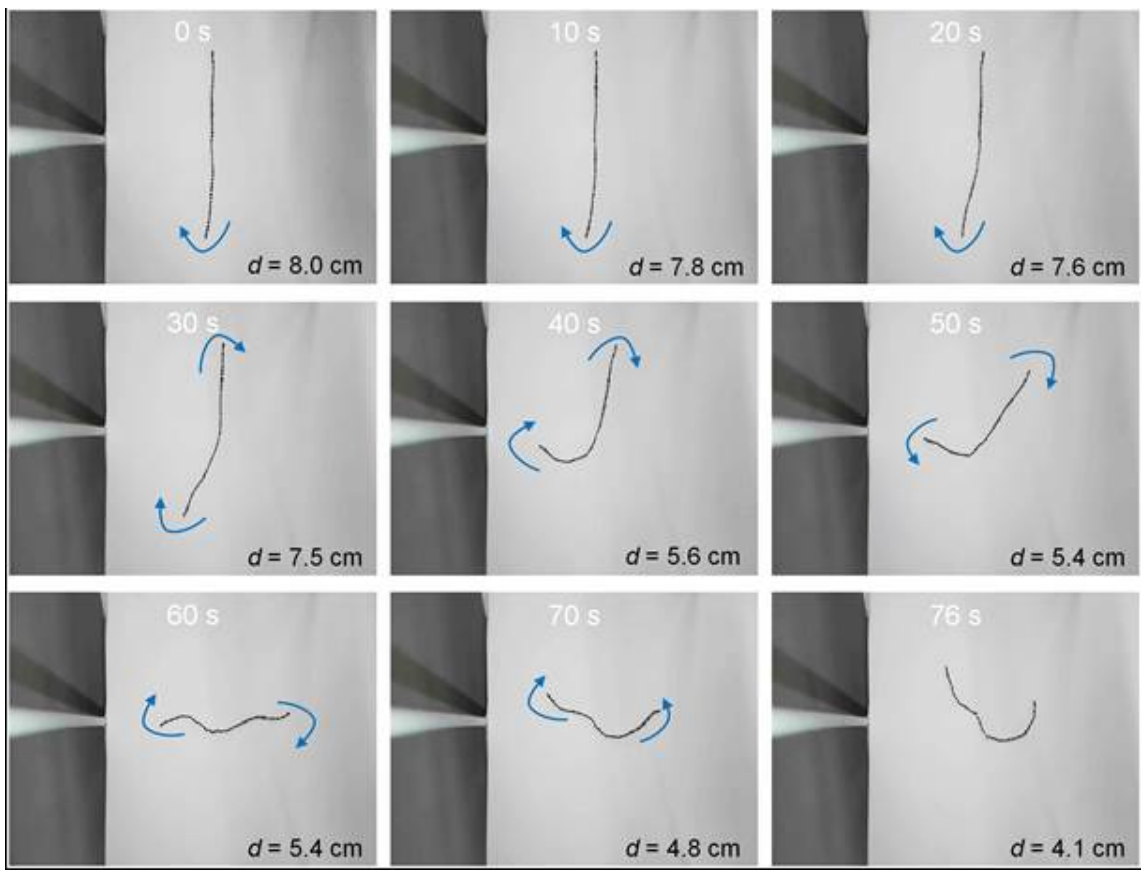

**Figure S13.** The motion of the two ends of a single twin-bundle actuator at different time of drying in an atmospheric environment (20 $^o$C, 70% RH) (from Movie S1). Note: the single twin-bundle actuator is placed on the surface of an A4 printing paper, and the blue arrows show the motion direction of the two ends.

As for the actuation mechanism, we suppose that it is caused by the free motion of the two ends of the single twin-bundle actuator. From the movie of the folding motion of the single twin-bundle actuator (Movie S1), we find that the two ends of the single twin-bundle actuator can easily change their places from time to time during the twisting. The folding actuation can be clearly observed by the decreasing distance ($d$, cm) of the two ends from 8.0 cm to 4.1 cm. This is actually different from the double twin-bundle actuator, triple twin-bundle actuator, and flat cloth actuator bearing 16 twin-bundles. The latter three cloth actuators contain more than one twin-bundle, and their two ends can not easily change their places upon actuation due to the restriction of other adjacent bundles. It should be noted that the motion of two ends can not be the same at any time, since there are some inevitable different factors such as the microstructure and the adsorbed water content of the single twin-bundle actuator. The motion of the single twin-bundle actuator is thus chaotic but very possibly not random, as its end-to-end distance can be quantitatively described mathematically by certain functions, similar to the theory used to describe the two ends of a polymer chain with a defined persistent length [S4].

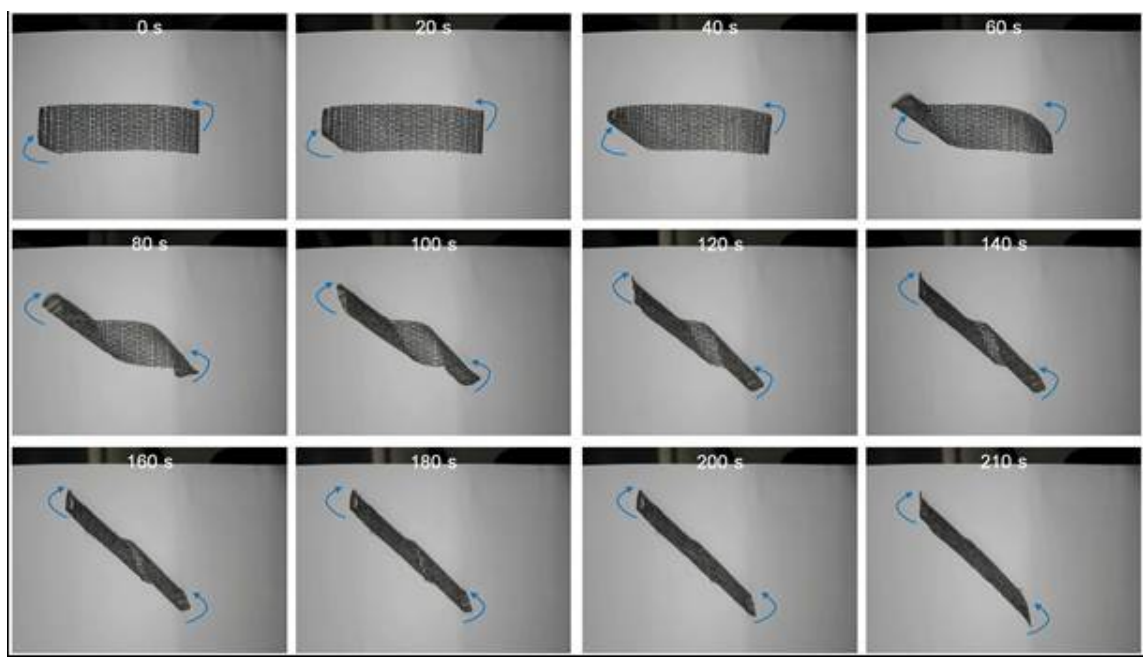

**Figure S14.** The motion of the flat cloth actuator bearing 16 twin-bundles at different time of drying in an atmospheric environment (20 $^{\circ}$C, 70% RH) (from Movie S3). Note: the flat cloth actuator is placed on the surface of a printing A4 paper, and the blue arrows show the motion direction of the two short sides.

In order to study the actuation mechanism, the rolling process of the flat cloth actuator is filmed (please see Movie S3). It is observed that the two short sides of the cloth actuator gradually roll to form a cylinder, due to the twisting of each bundle fibers.

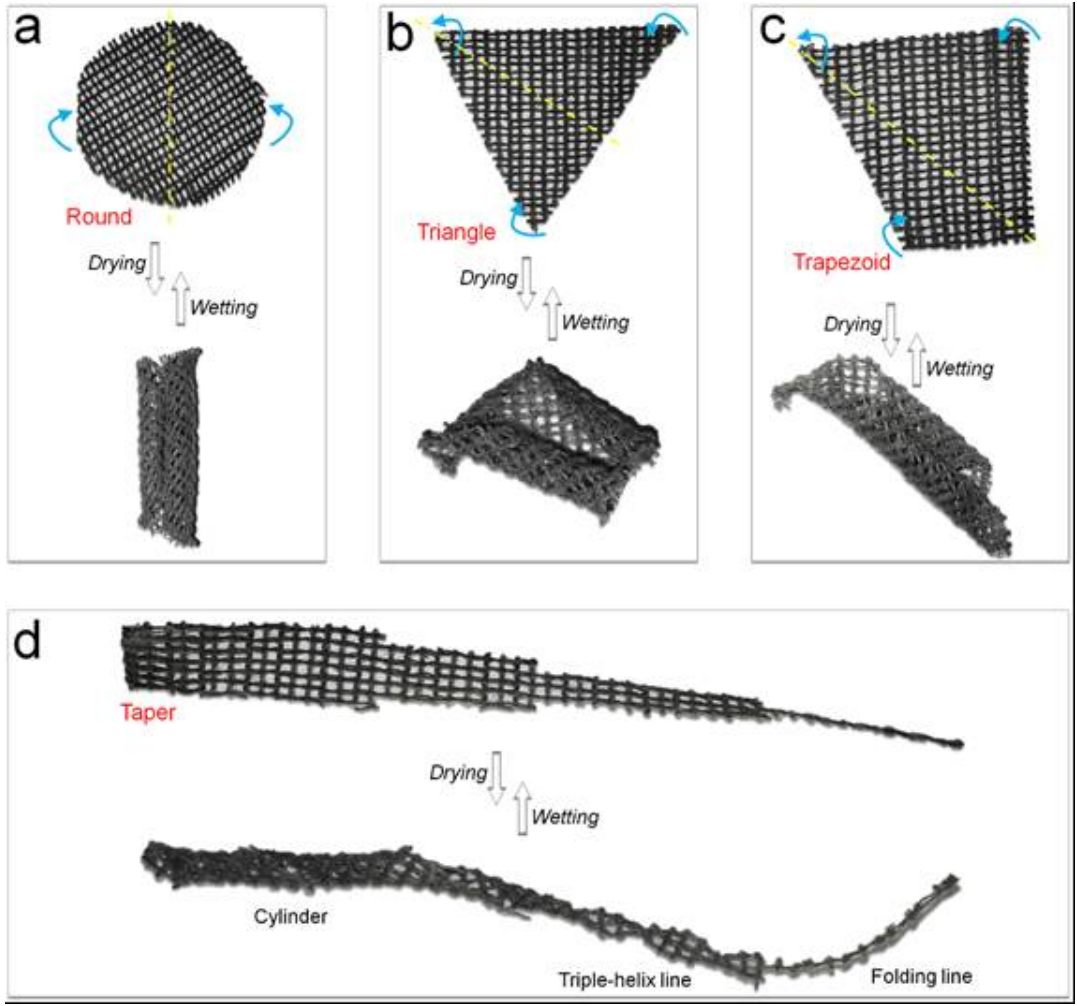

**Figure S15.** Photographs of cloth actuators with several more exotic shapes before and after actuation. (a) Round cloth actuator, (b) triangular cloth actuator, (c) trapezoid cloth actuator, and (d) taper cloth actuator bearing four parts with 7, 5, 3 and 1 fiber bundles, respectively. The blue arrows show the motion direction and the yellow dash lines show the axis direction.

Apart from the helix-line and cylinder shapes from the double twin-bundle actuator, triple twin-bundle actuator and flat cloth actuator as shown in Figure 3 in the main text, we prepare other kinds of cloth actuators including round, triangle and trapezoid to achieve several more exotic shapes. We even design a taper cloth actuator bearing four parts with 7, 5, 3 and 1 fiber bundles, respectively, to simultaneously achieve the folding line, helix line and cylinder. These results suggest that a broad variety of different shapes can be created from the cloth actuator that can deliver more complex actuation forms.

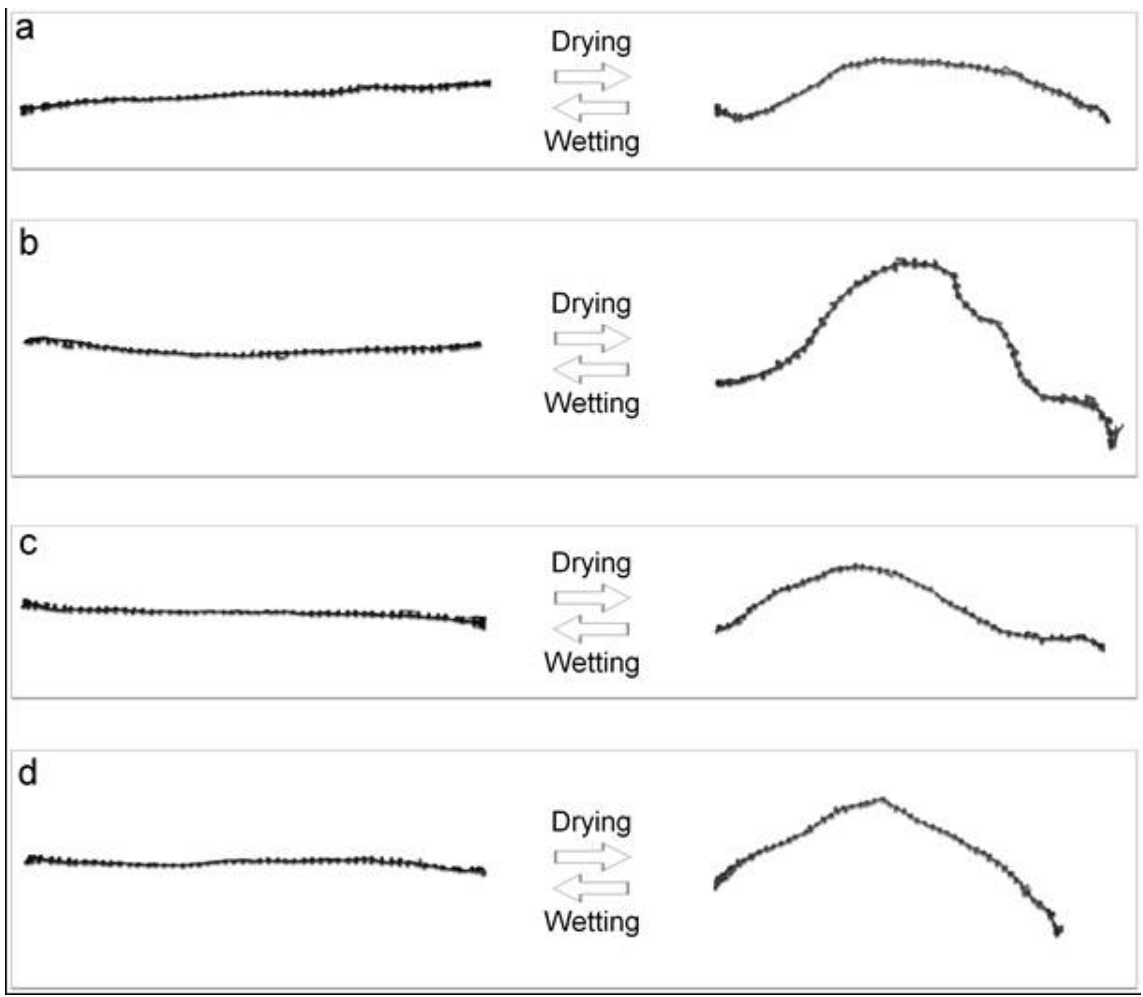

**Figure S16.** The reproducibility (a–d) of the actuation for the single twin-bundle actuator tested by four parallel samples.

The reproducibility of the actuation is confirmed by four parallel tests. Although none of them is identical due to the kinetical reason and/or intrinsic difference such as manufacturing, the macroscopic folding is similar to each other.

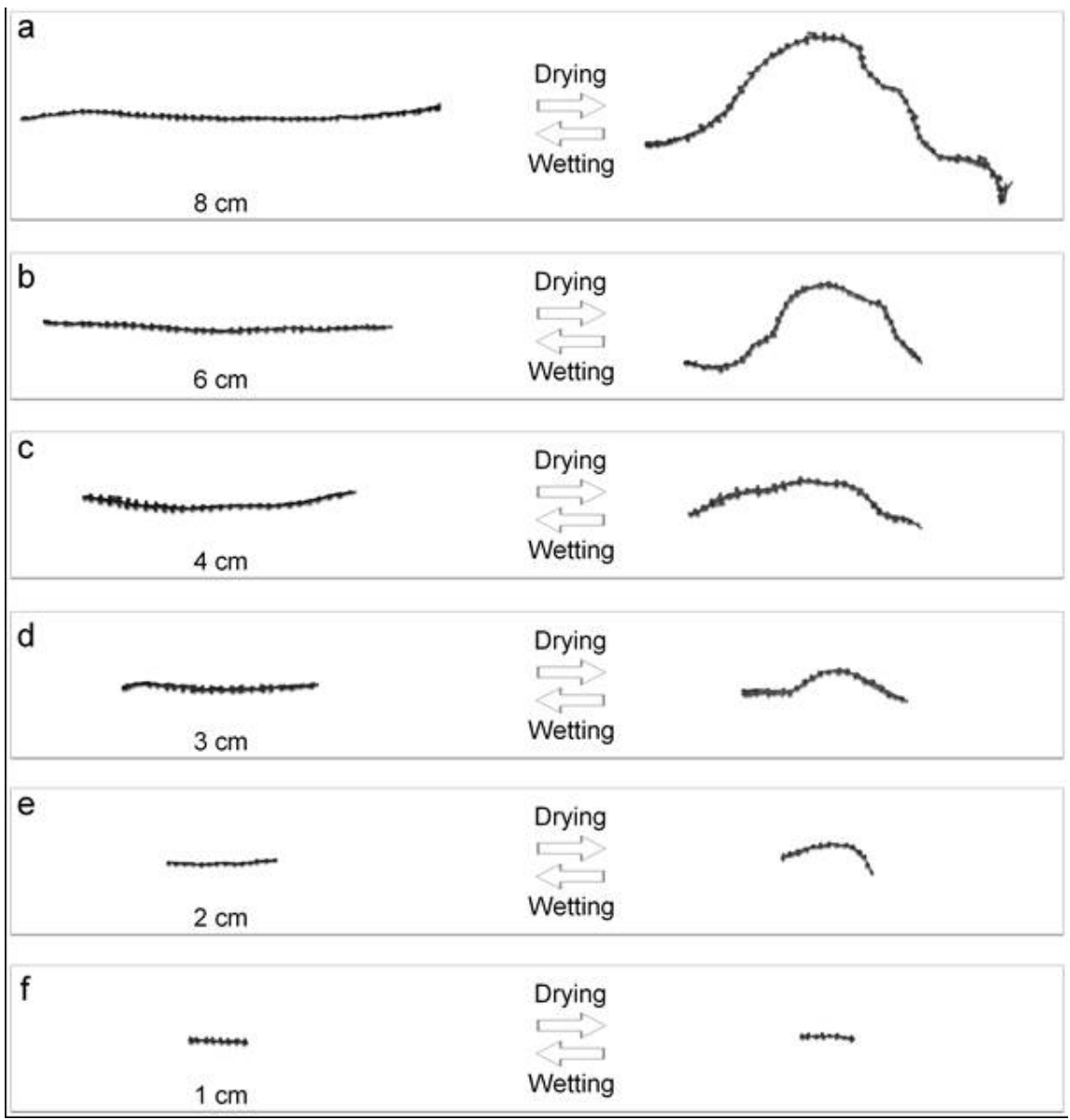

**Figure S17.** The effect of the length (a for 8 cm, b for 6 cm, c for 4 cm, d for 3 cm, e for 2 cm, and f for 1 cm) on the folding motion of the single twin-bundle actuator.

The actuation of the single twin-bundle actuator is modulatable in terms of its length. Evidently, the longer the single twin-bundle actuator is, the more obvious the line will be folded.

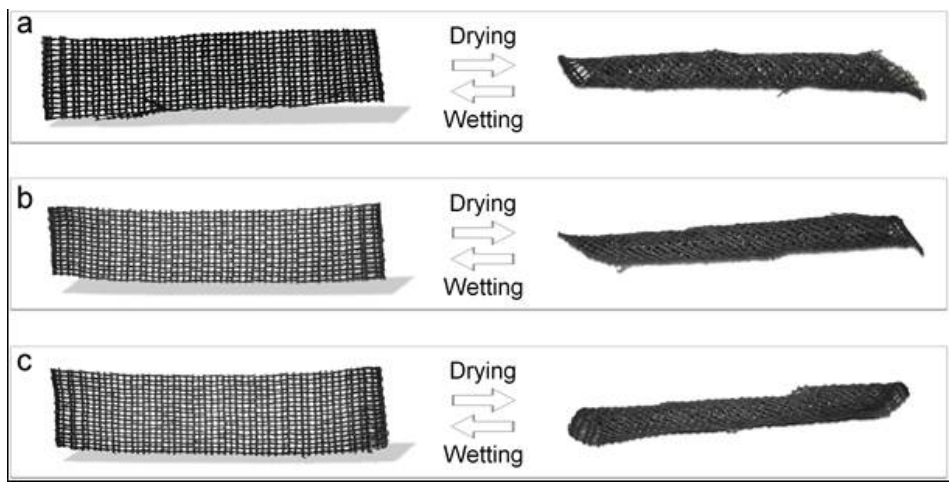

**Figure S18.** Photographs of some typical examples (a–c) for the reproducibility of the actuation of the flat cloth actuator through three parallel tests.

The reproducibility of the actuation for the flat cloth actuator was confirmed by at least ten parallel tests, and we find that all of them roll into a cylinder. The photographs of three typical samples are displayed.

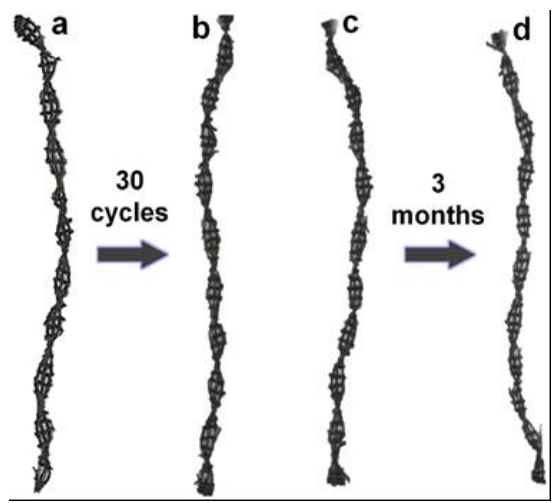

**Figure S19.** Photographs of the helical morphologies of the strip-like cloth actuator bearing triple twin-bundles after 1 cycle (a, 10 revolutions) and 30 cycles (b, 10 revolutions) of drying/wetting. Photographs of the helical morphologies of the strip-like cloth actuator before (c, 10 revolutions) and after storage in water for 3 months (d, 10 revolutions).

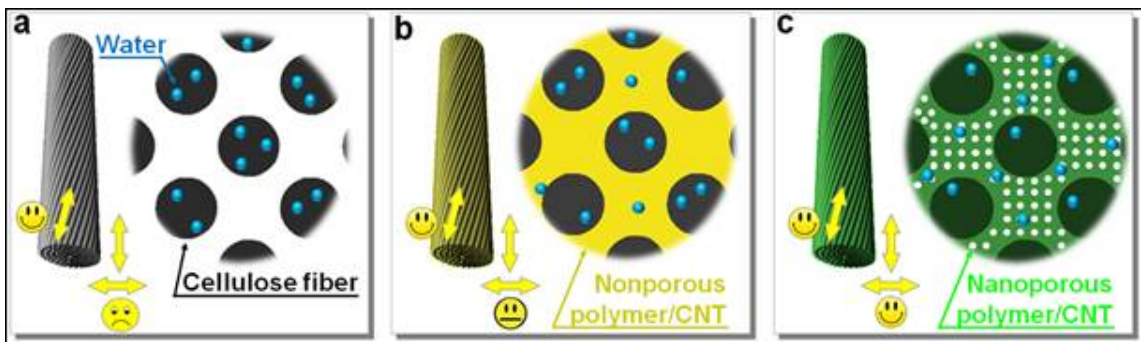

**Figure S20.** Schematic illustration of the enhanced actuation mechanism. Here only one bundle is taken as an example for simplicity; (a) for the native cloth, (b) for the nonporous actuator intermediate without ammonium treatment, and (c) for the cloth actuator bearing the nanoporous polymer/CNT hybrid; the yellow double arrows indicate the diffusion direction of water molecules in helically arranged microfibers.

The finite element simulation might be useful for the further understanding of the actuation mechanism for the cloth actuator. There are a lot of papers using the finite element simulation method [S5]. However, the difficulty in modeling the actuation behavior is due to the complex multilevel architecture of cloth actuator, and a detailed theoretical study of how the twisting actuation behavior operates at multiple length scales in our cloth actuator would be a study on its own, which is out of the scope of this manuscript. We intend to continue this work in the future from a theoretical perspective which we hope will provide more detailed insights about the actuation mechanism of the cloth actuator.

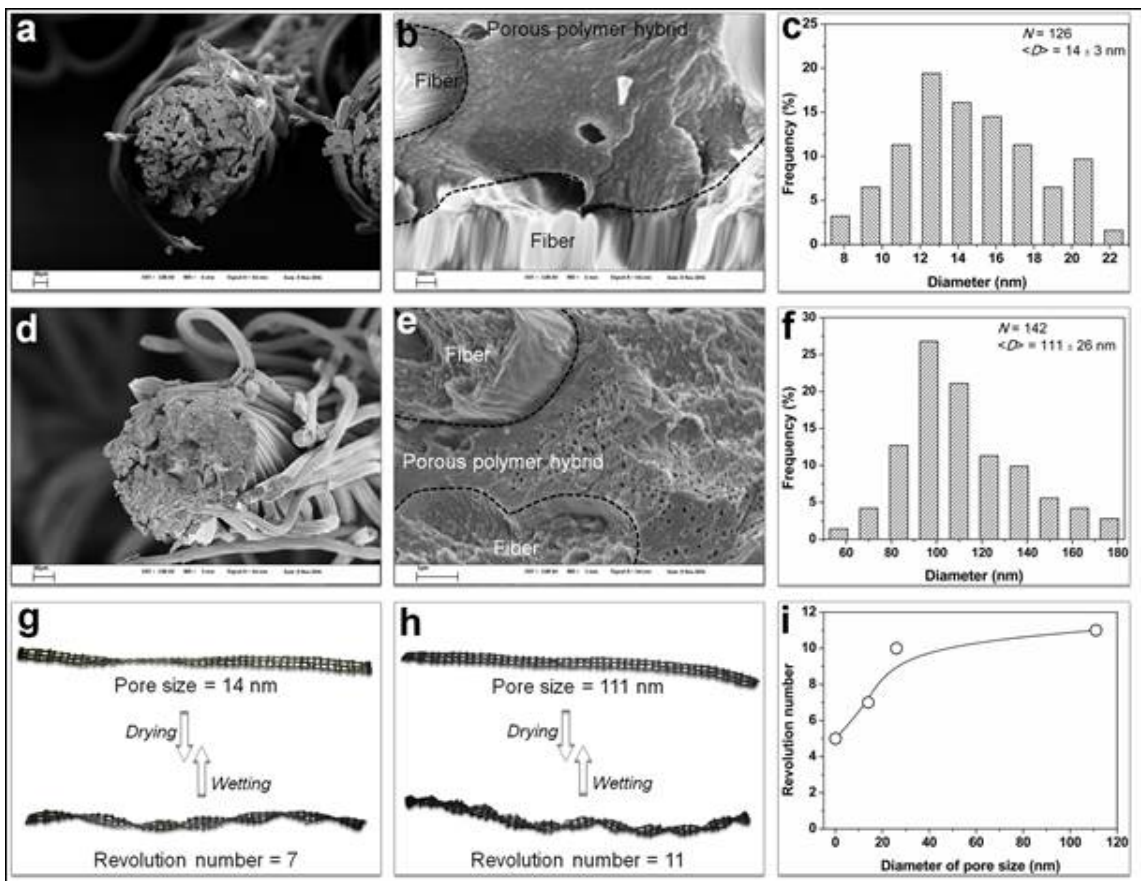

**Figure S21.** (a and b) SEM images of the cross-sectional structure and (c) the pore size distribution plot of the cloth actuator prepared by using PAA 450,000 g mol$^{-1}$. (d and e) SEM images of the cross-sectional structure and (f) the pore size distribution plot of the cloth actuator prepared by using isophthalic acid (IPA, 166 g mol$^{-1}$). (g) Photographs of the cloth actuator prepared by using PAA 450,000 g mol$^{-1}$ before and after actuation. (h) Photographs of the cloth actuator prepared by using IPA (166 g mol$^{-1}$) before and after actuation. (i) Plot of the revolution number of cloth actuators *vs.* the pore size in the porous polymer hybrid.

In order to vary the pore size of the porous PIL-PAA/CNT hybrid, the only practical method that allows us is to change the molecule weight of PAA. High molecule-weight PAA favors for the formation of the porous polymer hybrid with smaller pore size, while low molecule-weight PAA for bigger pore size.

Consequently, instead of PAA with the molecule weight of 1800 g mol$^{-1}$ (giving the pore size of 26 nm in the cloth actuator) in the main text, we also use PAA with the molecule weight up to 450,000 g mol$^{-1}$ to prepare the cloth actuator with a smaller pore size (14 nm). Compared to that of the cloth actuator prepared by using PAA 1800 g mol$^{-1}$, the revolution number of the cloth actuator prepared by using PAA 450,000 g mol$^{-1}$ is lower (7 *vs.* 10). This result suggests that the larger pores promote the actuation performance, possibly due to the enhanced diffusion kinetics of water among fibers.

Besides, since it is difficult to obtain commercial PAA with the molecule weight of less than 1800 g mol$^{-1}$ (*e.g.*, 100–200 g mol$^{-1}$), small organic acid was used instead. We here select IPA (166 g mol$^{-1}$) to access porous PIL-IPA/CNT hybrid with the pore size of 111 nm. We find that the pore size larger than 30 nm has little-to-no effect on the revolution number of the cloth actuator (11 *vs.* 10). These results above imply that the pore size effect is only visible when below 30 nm.

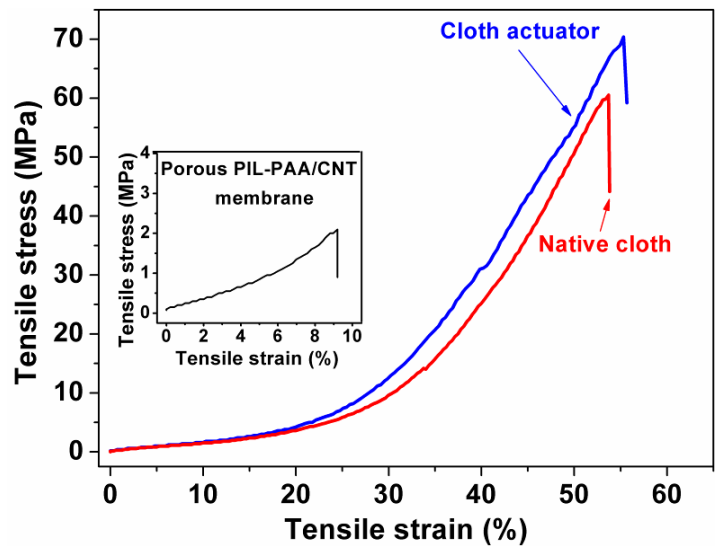

**Figure S22.** Tensile stress-strain curves of the cotton-free PIL-PAA/CNT porous membrane, the native cloth, and the resultant cloth actuator. The tensile stress is calculated *via* dividing the force (N) by the cross-sectional area of sample (m$^2$). The tensile strain is calculated *via* dividing the extension length of sample (m) by the original length (m).

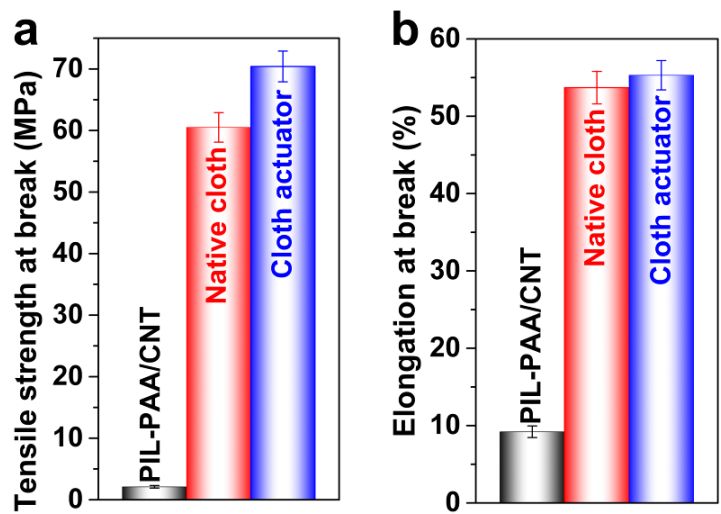

**Figure S23.** Comparisons of (a) tensile strength at break and (b) elongation at break of the cotton-free PIL-PAA/CNT porous membrane, the native cloth, and the resultant cloth actuator. All data were the average of five independent measurements.

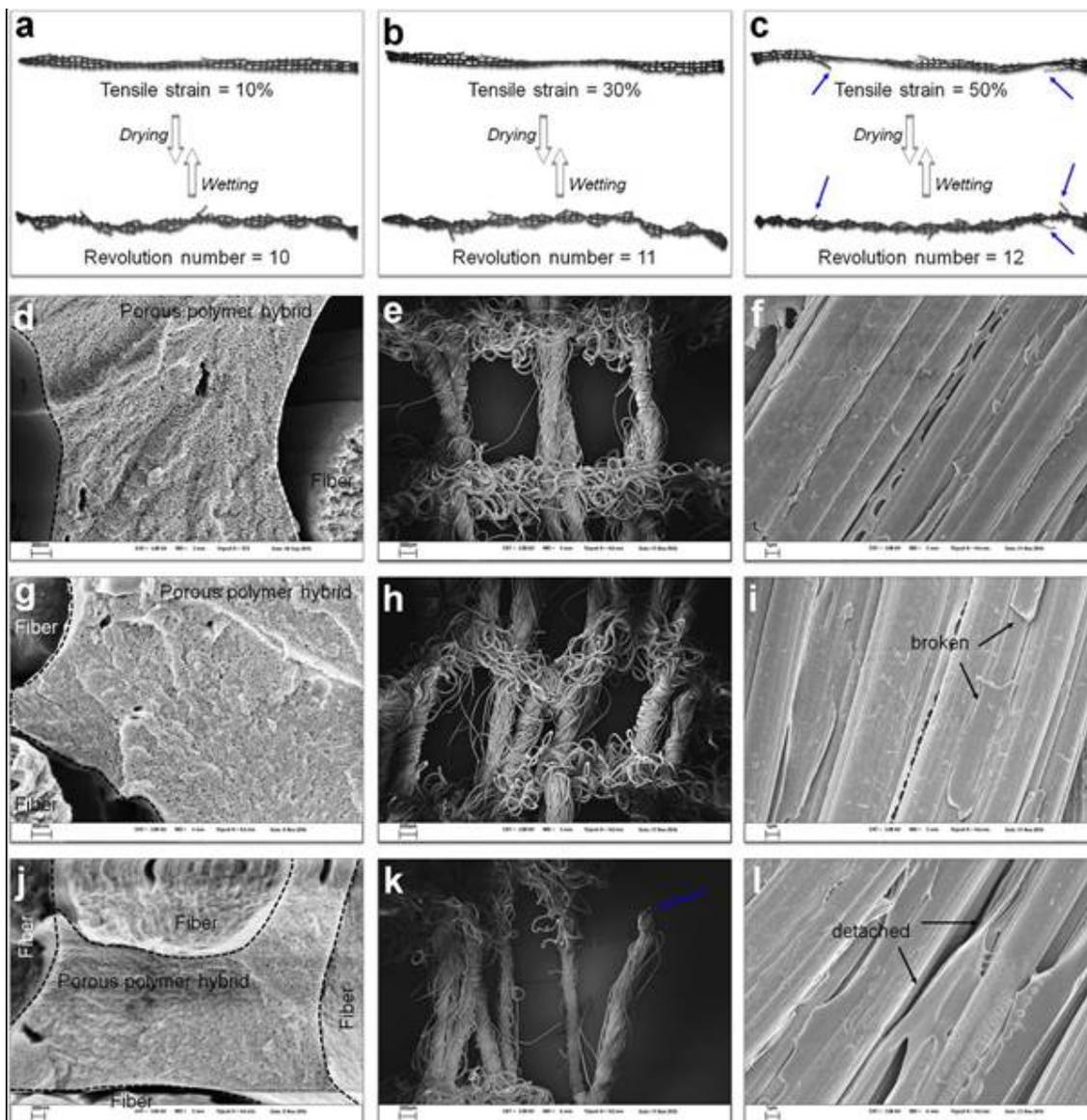

**Figure S24**. Photographs of the pre-treated cloth actuators with the tensile strain of (a) 10%, (b) 30% and (c) 50%, respectively, before and after actuation. SEM images of the pre-treated cloth actuators with the tensile strain of 10% (d for the cross-sectional view; e and f for the front view), 30% (g for the cross-sectional view; h and i for the front view), and 50% (j for the cross-sectional view; k and l for the front view). Note: the blue arrows in (c) and (k) show the broken part of some bundles, and the black arrows in (i) and (l) show the broken and detached parts of the porous polymer hybrid on the surface of fibers.

In order to study whether the cloth actuator still works properly after the application of large tensile strain and explore the effect of the tensile strain on the actuation performance of the cloth actuator, the freshly prepared cloth actuator is firstly stretched to yield a tensile strain of 10%, 30% and 50%, respectively, considering that the elongation at break of the cloth actuator is ca. 55.3%.

As we can see, when the pre-treated tensile strain is 10%, the revolution number of the pre-treated cloth actuator (10) is identical to that of the untreated one. Surprisingly, when the pre-treated tensile strain increases to 30% and 50%, the revolution number of the pre-treated cloth actuator slightly goes up to 11 and 12, respectively.

Since the pore size of the porous PIL-PAA/CNT hybrid in the pre-treated cloth actuators does not change (25–30 nm, as shown in (d), (g) and (j)) compared to that of the freshly prepared cloth actuator (26 nm), the possible reason for the increasing revolution number at 30% and 50% tensile strain is the elongation of the cloth actuator. We find that the length of the cloth actuator increases from 8.0 cm to 8.5 and 9.0 cm when the pre-treatment of tensile strain goes up to 30% and 50%, respectively. It should be pointed out that although the larger tensile strain yields higher revolution number, a small fraction of the porous polymer hybrid on the surface of fibers and a part of the fiber bundles are broken up, which is actually unfavorable for the practical applications of the cloth actuator, particularly for the long-term repeated uses. Therefore, the suggested maximum tensile strain of the cloth actuator is 30%.

Nevertheless, the above results demonstrate that the cloth actuator still works well upon moderate tensile strain up to 30%.

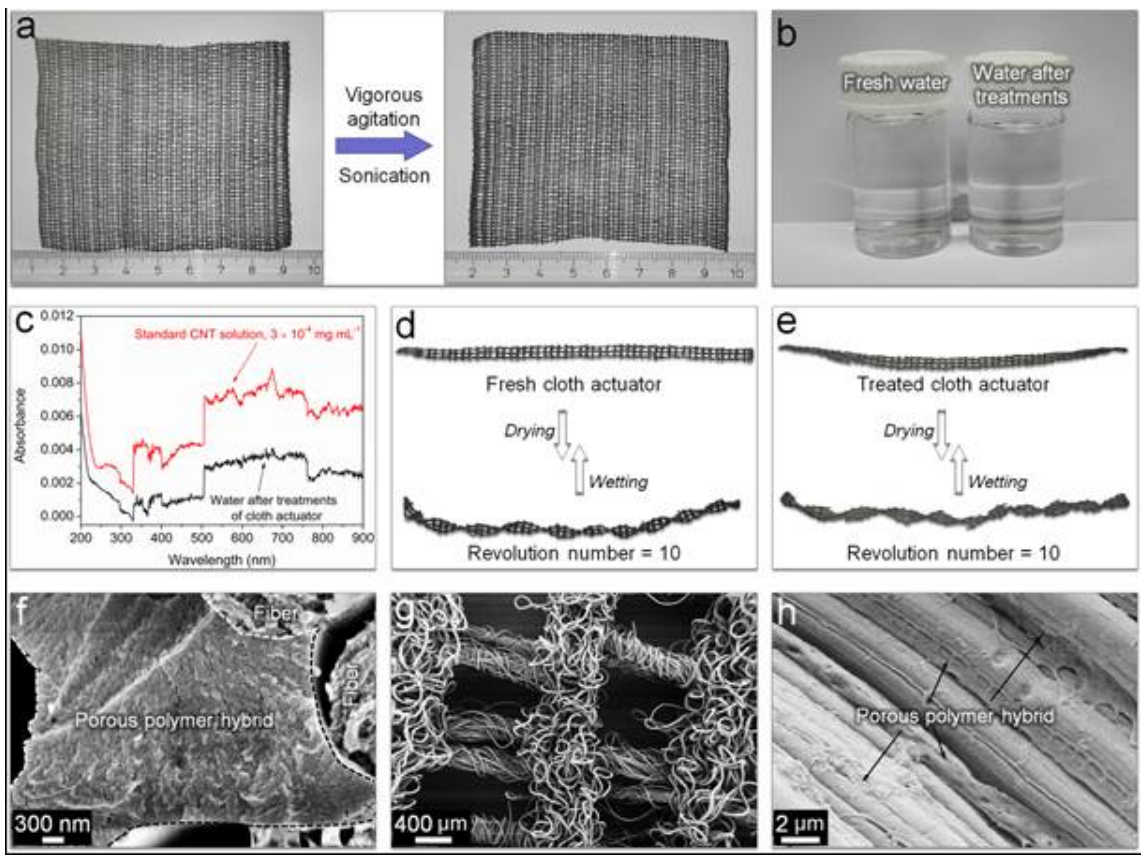

**Figure S25.** Robustness of the cloth actuator under washing condition. (a) Photographs of the cloth actuator before (left) and after (right) treatment. (b) Photographs of (left) the fresh water and (right) the water residue after treatment. (c) UV/Vis absorbance of the CNT aqueous solution with a concentration of $3 \times 10^{-4}$ mg mL$^{-1}$ and the water residue after treatment. The adaptive movements of the cloth actuator before (d) and after (e) treatment. SEM images of the treated cloth actuator are shown in (f) for cross-sectional view, and (g) and (h) for front view.

Experimental condition: The fresh cloth actuator (size = 8.0 cm × 6.7 cm) was washed in 200 mL water at an agitation rate of 1400 rpm (Asynt ltd., model ADS-HP1) for 3 h and then by sonication (Sandelin Sonnorex RK100H, 320 W) for another 3 h. Photographs of the cloth actuator before and after treatment are displayed in (a). Macroscopically no difference can be told. The solution in which the cloth actuator was soaked was photographed before and after treatment and is shown in (b). The solution after treatment remains transparent. *Via* UV/Vis spectroscopy, we investigated this solution and did detect a weak signal of CNTs (as shown in (c)), which after calibration is determined to be below $3 \times 10^{-4}$ mg mL$^{-1}$ and thus negligible regarding the CNTs incorporated into the actuator (< 0.06 mg *vs.* 2.37 mg, that is to say, < 3 wt %).

Furthermore, the imidazolium cation of the PIL has a characteristic absorbance band at 400 nm [S6], while the UV/Vis spectrum in (c) did not detect this band, which thereby excludes the presence of PIL in solution. Besides, upon drying, the treated cloth actuator bearing triple twin-bundles exhibits a revolution of 10, identical to the fresh cloth actuator (as displayed in (d) and (e)). SEM images of the cross-section and the surface of the treated cloth actuator are shown in (f–h), which are similar to the non-treated sample (Figure 2 in the manuscript), thus confirming the structural integrity of the porous PIL hybrid networks. These results prove that the cloth actuator is stable against washing with little-to-no leakage of structural component.

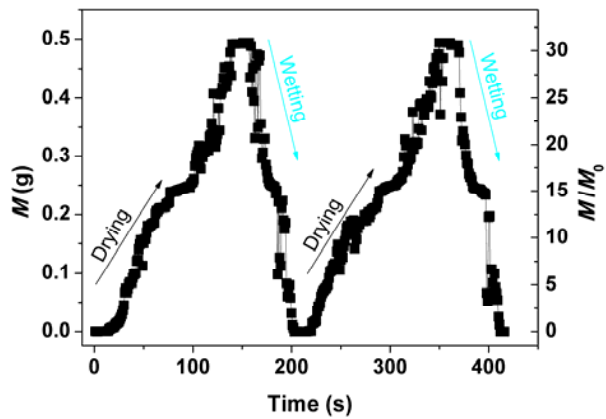

**Figure S26**. The force *vs.* time plot of the native cloth bearing triple twin-bundles during rotation. Note: *M* in the form of weight is the force exerted on the balance by the native cloth; $M_0$ is the weight of the native cloth.

The maximum force detected in the native cloth is 0.49 g, which is only one-fourth of that of the cloth actuator (2.2 g).

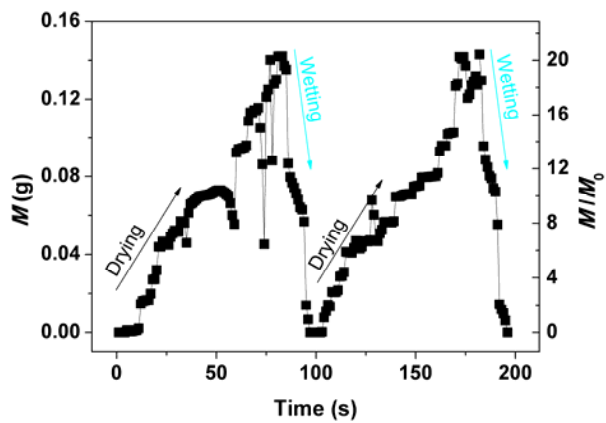

**Figure S27**. The force *vs.* time plot of a single twin-bundle actuator. Note: *M* in the form of weight is the force exerted on the balance by the single twin-bundle actuator; $M_0$ is the weight of the single twin-bundle actuator.

The maximum force detected in the single twin-bundle actuator is 0.14 g, which is 20 times that of its own weight (0.007 g).

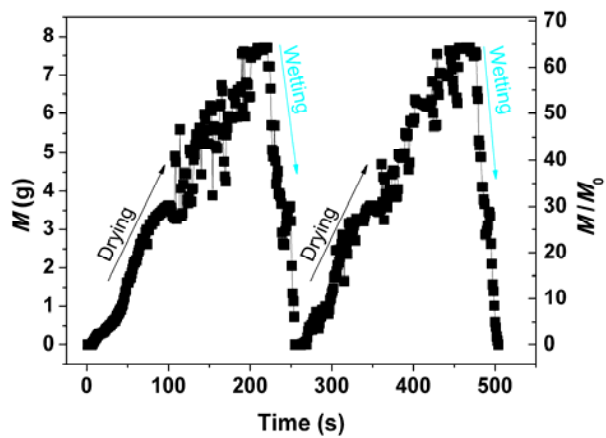

**Figure S28.** The force *vs.* time plot of the flat cloth actuator. Note: *M* in the form of weight is the force exerted on the balance by the flat cloth actuator; $M_0$ is the weight of the flat cloth actuator. The maximum force detected in the flat cloth actuator is 7.70 g, which is ca. 63 times that of its own weight (0.120 g).

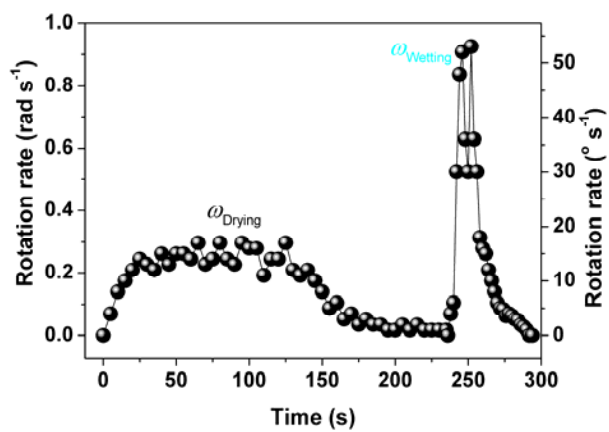

**Figure S29.** The rotation rate *vs.* time plot of the native cloth bearing triple twin-bundles.

The maximum rotation rate of the native cloth is ca. 0.9 rad s$^{-1}$. The initial reference acceleration ($\alpha$) upon wetting is calculated to be 0.04 rad s$^{-2}$, which is only one-seventh of that of the cloth actuator (0.28 rad s$^{-2}$). Since the moment of inertia of the reference (*J*) is 4.5 × 10$^{-7}$ kg m$^{-2}$, the maximum start-up torque ($\tau$) is $\tau = J\alpha = 0.18 \times 10^{-7}$ N m$^{-1}$, also one-seventh of that of the cloth actuator (1.26 × 10$^{-7}$ N m$^{-1}$). Thereby, the introduced nanoporous polymer hybrid significantly enhances the mechanical performance of the cloth actuator.

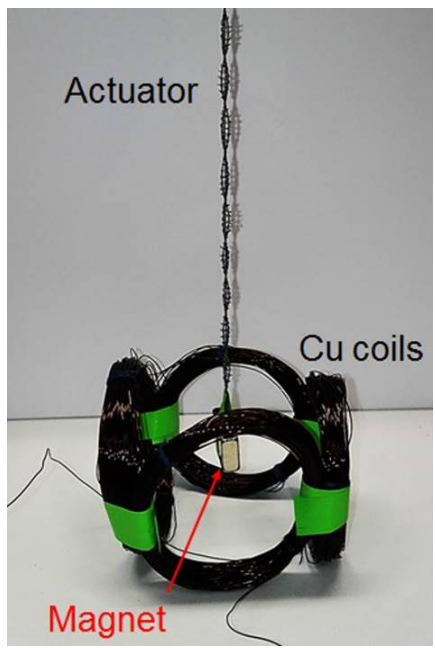

**Figure S30.** Photograph of the setup for the electric generator made of the strip-like cloth actuator bearing triple twin-bundles (3.8 mm × 160 mm), a magnet, and dense copper coils.

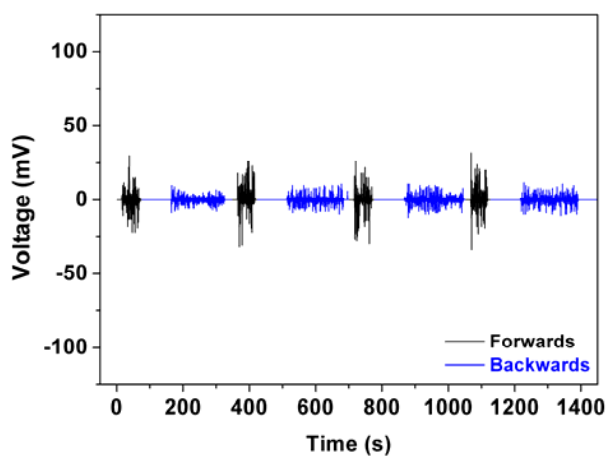

**Figure S31.** Plot of the open-circuit voltage produced by the rotation (forwards by wetting and backwards by drying) of the native cloth bearing triple twin-bundles *vs.* time.

The yielded open-circuit voltage produced by the native cloth (~24 mV) is only one-third of that of the cloth actuator (75 mV). As a result, the cloth actuator is superior to the native cloth, justifying the key importance of the porous polymer hybrid incorporated in the cloth actuator.

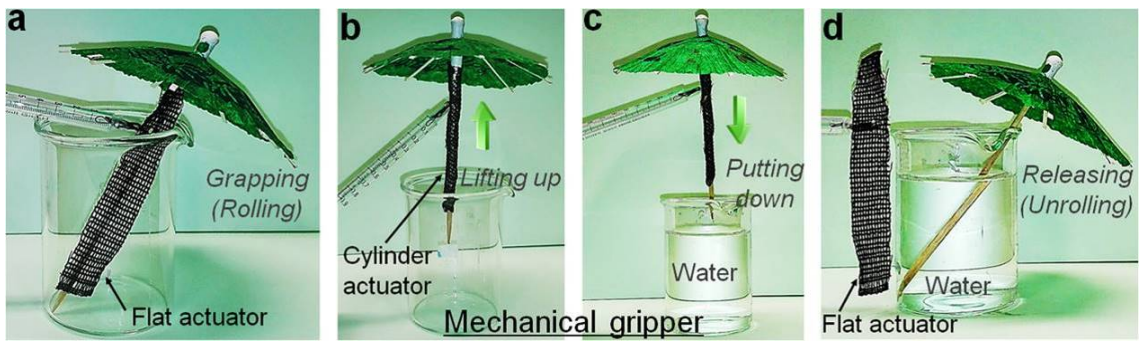

**Figure S32.** The mechanical gripper made of the flat cloth actuator containing 16 twin-bundles (20 mm × 80 mm); (a) the grapping of a small umbrella by the rolling of the cloth actuator, (b) the lifting-up of umbrella, (c) the dropping of actuator-surrounded umbrella into water, and (d) the release of umbrella from the cloth actuator by the unrolling of the cloth actuator.

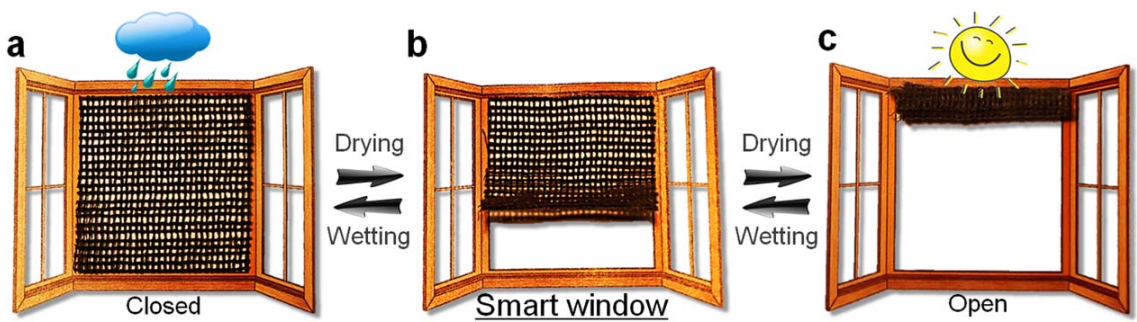

**Figure S33.** The "smart" window made of the cloth actuator (32 mm × 34 mm); (a) the "closed window" (flat actuator) in rainy days, (b) the recovery, and (c) the "open window" (cylinder actuator) in sunny days.

**References to SI**